\documentclass[aps,prl,twocolumn,superscriptaddress]{revtex4-1}

\usepackage[usenames,dvipsnames,svgnames,table]{xcolor}
\usepackage{graphicx}
\usepackage{amsmath}

\begin{document}

\title{Coherence Times of Bose-Einstein Condensates beyond the Shot-Noise Limit via Superfluid Shielding}

\author{William Cody Burton}
\affiliation{Research Laboratory of Electronics, MIT-Harvard Center for Ultracold Atoms, Department of Physics, Massachusetts Institute of Technology, Cambridge, Massachusetts 02139, USA}

\author{Colin J. Kennedy}
\affiliation{Research Laboratory of Electronics, MIT-Harvard Center for Ultracold Atoms, Department of Physics, Massachusetts Institute of Technology, Cambridge, Massachusetts 02139, USA}

\author{Woo Chang Chung}
\affiliation{Research Laboratory of Electronics, MIT-Harvard Center for Ultracold Atoms, Department of Physics, Massachusetts Institute of Technology, Cambridge, Massachusetts 02139, USA}

\author{Samarth Vadia}
\affiliation{Research Laboratory of Electronics, MIT-Harvard Center for Ultracold Atoms, Department of Physics, Massachusetts Institute of Technology, Cambridge, Massachusetts 02139, USA}
\affiliation{Fakult\"{a}t f\"{u}r Physik, Ludwig-Maximilians-Universit\"{a}t, Schellingstr. 4, 80799 M\"{u}nchen, Germany}{

\author{Wenlan Chen}
\affiliation{Research Laboratory of Electronics, MIT-Harvard Center for Ultracold Atoms, Department of Physics, Massachusetts Institute of Technology, Cambridge, Massachusetts 02139, USA}

\author{Wolfgang Ketterle}
\affiliation{Research Laboratory of Electronics, MIT-Harvard Center for Ultracold Atoms, Department of Physics, Massachusetts Institute of Technology, Cambridge, Massachusetts 02139, USA}

\date{26 October 2016}

\begin{abstract}
We demonstrate a new way to extend the coherence time of separated Bose-Einstein condensates that involves immersion into a superfluid bath. When both the system and the bath have similar scattering lengths, immersion in a superfluid bath cancels out inhomogeneous potentials either imposed by external fields or inherent in density fluctuations due to atomic shot noise. This effect, which we call superfluid shielding, allows for coherence lifetimes beyond the projection noise limit.  We probe the coherence between separated condensates in different sites of an optical lattice by monitoring the contrast and decay of Bloch oscillations. Our technique demonstrates a new way that interactions can improve the performance of quantum devices.
\end{abstract}

\maketitle

Phase coherence between spatially separated quantum objects is a central theme of quantum physics, with direct relevance for many applications in quantum information, quantum simulation \cite{ketterle2013a,bloch2013,ketterle2015}, atom interferometry \cite{pritchard2009}, and force sensing \cite{ignuscio2005,tino2006,stamperkurn2014}. Quantum mechanics fundamentally limits the fidelity with which one can split a coherent object, perform an operation on the separated parts of the system, and read out phase information via interference.  Often, the coherence time is limited by technical fluctuations or by interactions with the environment. For non-interacting systems, classical shot noise determines the signal-to-noise ratio and the final precision in the measurement of the phase. The relative uncertainty scales with the number of events $N$ as $1/\sqrt{N}$, and coherence time is independent of $N$.  In an interacting system, on the other hand, the coherence time is often set by shot noise, as number fluctuations cause fluctuations of the chemical potential: $\delta \mu = \delta N \times \left|\partial \mu/\partial N\right|$. Modifying interactions can change $\left|\partial \mu/\partial N \right|$, which can lead to a long coherence time \cite{naegerl2008}. Another way of improving the limitations set by shot noise is squeezing the uncertainty by using nonlinear interactions between modes of the system \cite{vuletic2010} or by using quantum measurements \cite{thompson2014}. Highly spin-squeezed states have been observed for pairs \cite{ketterle2007,ketterle2007b} and arrays \cite{oberthaler2008} of independent Bose-Einstein condensates (BECs). 

In this Letter, we present a new method of enhancing the phase coherence time of separated BECs beyond the shot noise limit by immersing the system into a superfluid bath---an effect we will call superfluid shielding.  The superfluid bath, through its interactions with the system, compensates for technical and number fluctuations which would otherwise shorten the coherence time.  In an optical lattice, the phase coherence and number fluctuations of a BEC can be probed by the time evolution after a rapid projection of the state onto a localized basis, either through a fast ramp to high lattice depths \cite{bloch2010}, or by the sudden application of a large acceleration to the lattice. The second case leads to the phenomenon of Bloch oscillations, which we use to create separated condensates and probe their phase coherence. By tracking the evolution of Bloch oscillations, we demonstrate that superfluid shielding can shield inhomogeneities created both by external fields common to both spin species (e.g. an optical trapping potential) and by fundamental projection noise. A theoretical analysis shows that fluctuations in the chemical potential can be reduced by up to two orders of magnitude for $^{87}$Rb condensates.

\begin{figure}
\includegraphics[width=\columnwidth]{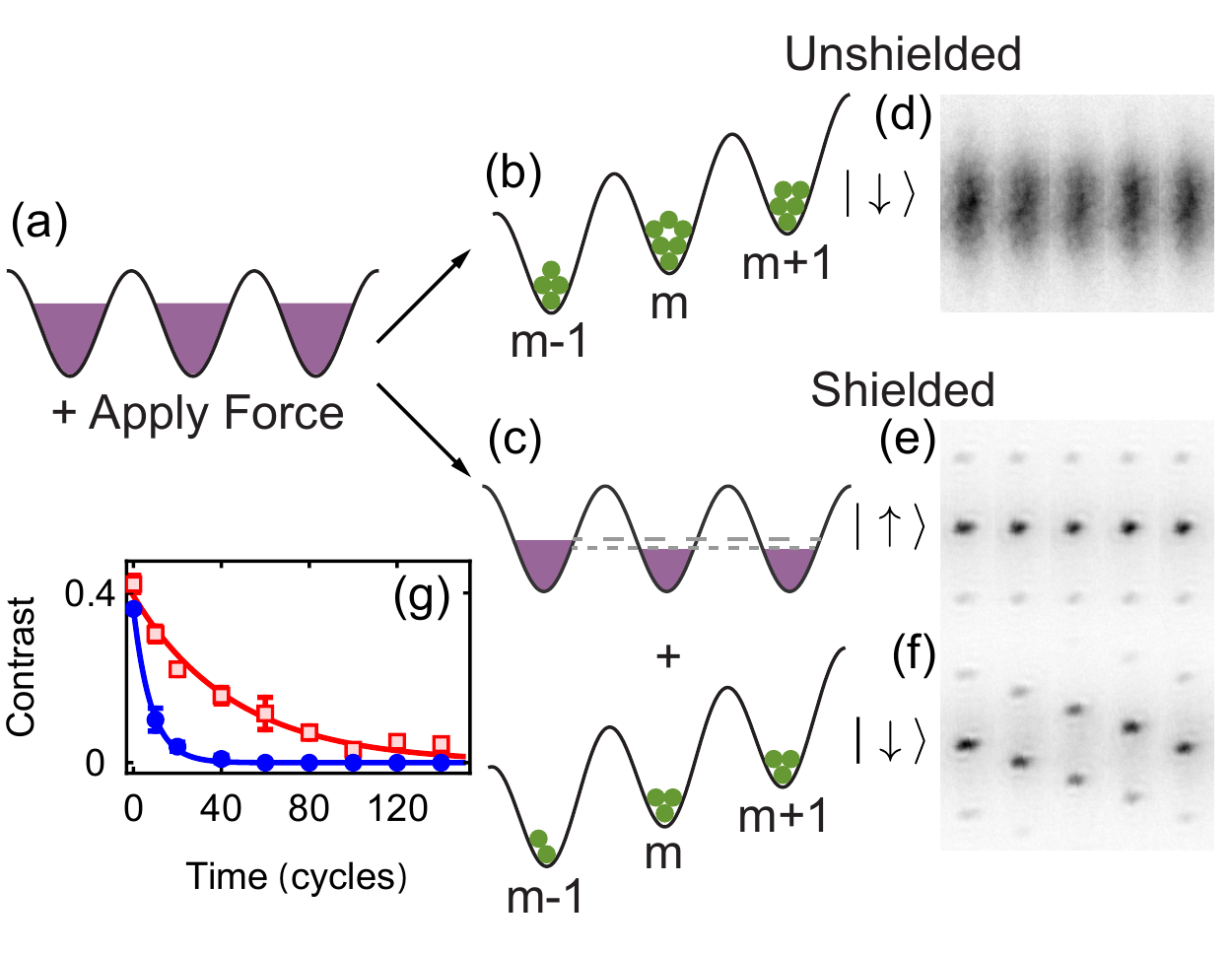}
\caption{\label{Fig1}\label{Fig2}(Color online) Schematic of superfluid shielding. (a) Before applying a tilt, the atoms are in a superfluid, which is approximately described by a coherent state on each site. The chemical potential is constant across the cloud. (b) In the limit of a strong tilt ($\Delta \gg J$), the wavefunction at each lattice site is projected onto the number basis, leading to fluctuations in the number of atoms and chemical potential from site to site. (c) If the gas has two components, one which is localized by the tilt, and one which remains superfluid, the itinerant component can compensate for fluctuations in the localized component. (d-f) Momentum distribution over the course of a single Bloch oscillation after ten cycles. (d) Without superfluid shielding, the diffuse cloud indicates decoherence of the condensate. (e) The itinerant component feels no force and does not Bloch oscillate. (f)  For the shielded component, the Bloch oscillation contrast is high. (g) Exponential decay of the Bloch oscillation contrast for a one-component (blue dots) and two component (red squares) gas, for a transverse lattice depth of  $11\;E_r$ and $\sim$$8 \times 10^3$ atoms.}
\end{figure}

We begin with a BEC in an optical lattice, and create separated BECs by suddenly applying a strong potential gradient. The tilted potential suppresses resonant tunneling and allows the now separated condensates to evolve independently.  In this regime, the energy offset between adjacent lattice sites $\Delta$ is much larger than the bandwidths $\sim$$4J$, where $J$ is the nearest-neighbor tunneling matrix element. Most previous studies on Bloch oscillations have been done in the low-tilt regime \cite{korsch2005, naegerl2014, kolovsky2015a} where adjacent Wannier-Stark states overlap.  It has been shown in this regime that the presence of quasi-random disorder causes dephasing of Bloch oscillations that can be partially compensated by weak interactions \cite{schneble2010, santos2008} if the disorder is present before the tilt. Our work addresses the very different situation of a two-component system, where the second component does not feel the applied force and remains free to shield fluctuations created during or after the tilt, including fluctuations in the chemical potentials of the separated condensates due to projection noise.

A graphical description of superfluid shielding is presented in Fig. \ref{Fig1}a-c.  After the sudden tilting of the $\left|\downarrow\right>$ atoms, without shielding, the precession of the phase on each site is given by the local chemical potential and its fluctuations, $\mu^{\downarrow}_{j}$. When immersed in the superfluid of $\left|\uparrow\right>$ atoms, which are not subject to the tilt, the $\left|\downarrow\right>$ atoms experience repulsive interactions with the $\left|\uparrow\right>$ component. Because $\left|\uparrow\right>$ atoms remain itinerant, these atoms can freely adjust their local density in order to counteract the fluctuations in $\left|\downarrow\right>$ atoms and thus maintain a uniform global chemical potential.  Fluctuations in the density of the $\left|\downarrow\right>$ atoms, which would normally lead to chemical potential fluctuations and dephasing, are now anti-correlated with the $\left|\uparrow\right>$ density.

A more quantitative understanding of the system's response to both inhomogenous potentials $\delta V$ and density fluctuations $\delta n^\downarrow$ can be developed by examining the two-component Gross-Pitaevskii equation in the Thomas-Fermi approximation given by:
\begin{eqnarray}
\mu^{\downarrow}_{j} = g^{\downarrow\downarrow}n^{\downarrow}_{j}(x,y)+V^{\downarrow}_{\text{ext},j}(x,y)+g^{\uparrow\downarrow}n^{\uparrow}_{j}(x,y) \\
\mu^{\uparrow}_{j} = g^{\uparrow\uparrow}n^{\uparrow}_{j}(x,y)+V^{\uparrow}_{\text{ext},j}(x,y)+g^{\uparrow\downarrow}n^{\downarrow}_{j}(x,y),
\end{eqnarray}
where $\mu^{(\uparrow,\downarrow)}_j$ and $n^{(\uparrow,\downarrow)}_j$ are the chemical potential and number density, respectively, for a given component and site index $j$. The interaction terms $g^{\uparrow \uparrow}$, $g^{\downarrow \downarrow}$, and $g^{\uparrow \downarrow}$ are given by $4 \pi \hbar^2 a^{(\uparrow \uparrow,\downarrow \downarrow, \uparrow \downarrow)}/m$ where $a^{(\uparrow \uparrow,\downarrow \downarrow, \uparrow \downarrow)}$ are the $s$-wave scattering lengths for intra- and inter-component collisions. Before applying the tilt, $V^\downarrow_{\text{ext},j}$ and $V^\uparrow_{\text{ext},j}$ are both given by a common-mode harmonic trapping potential $V_{\text{trap},j}$, and since both components are superfluid, $\mu^{\left(\uparrow,\downarrow\right)}_j$ are constant across the cloud and independent of $j$. For a single-component system, this implies that the trapping potential is fully compensated by the inhomogeneous Thomas-Fermi density profile \cite{naegerl2010}.

When a spin dependent tilt of $\Delta$ per lattice site is applied, number fluctuations in the $\left|\downarrow\right>$ component, $\delta n^\downarrow_j$, are frozen in. We also allow for spin-independent potential fluctuations, so the total potentials are now $V^{\uparrow}_{\text{ext},j} = V_{\text{trap},j} + \delta V_j$ and $V^{\downarrow}_{\text{ext},j} = V_{\text{trap},j} + \delta V_j - j\Delta$. The $\left|\uparrow\right>$ component remains superfluid and therefore keeps a constant chemical potential $\mu^\uparrow_j = \mu^\uparrow$. This gives density fluctuations that are anti-correlated with both $\delta n^\downarrow_j$ and $\delta V_j$:
\begin{equation}
\delta n^\uparrow_j = -\frac{\delta V_j}{g^{\uparrow \uparrow}}-\frac{g^{\uparrow \downarrow}}{g^{\uparrow \uparrow}} \delta n^\downarrow_j.
\end{equation}
The back-action of $\delta n^\uparrow_j$ on the chemical potential for the $\left|\downarrow\right>$ component,
\begin{equation}
\begin{aligned}
\mu^{\downarrow}_{j} & = g^{\downarrow\downarrow}\left(n^{\downarrow}_{j}+\delta n^\downarrow_j\right)+V^{\downarrow}_{\text{ext},j}+g^{\uparrow\downarrow}\left(n^{\uparrow}_{j}+\delta n^\uparrow_j\right) \\
    & = \mu^{\downarrow,0} - j\Delta + \delta\mu^\downarrow_j
\end{aligned}
\end{equation}
where $\mu^{\downarrow,0}$ is the constant chemical potential of $\left|\downarrow\right>$ before the tilt is applied,
leads to a reduction in the fluctuations
\begin{equation} \label{shieldedMu}
\delta \mu^\downarrow_j = \eta_1 g^{\downarrow \downarrow} \delta n^\downarrow_j + \eta_2 \delta V_j,
\end{equation}
by the factors $\eta_1 = \left(g^{\uparrow \uparrow} g^{\downarrow \downarrow}-\left(g^{\uparrow \downarrow}\right)^2\right)/g^{\uparrow \uparrow} g^{\downarrow \downarrow}$  and $\eta_2 = \left(g^{\uparrow \uparrow}-g^{\uparrow \downarrow}\right)/g^{\uparrow \uparrow}$, which are both small for $g^{\uparrow\uparrow}\approx g^{\downarrow\downarrow} \approx g^{\uparrow\downarrow}$. Then the chemical potential is nearly independent of both common-mode potential fluctuations and atom number fluctuations, and depends only on the state-specific potential $j\Delta$, leading to long-lived Bloch oscillations. In $^{87}$Rb, all scattering lengths between hyperfine ground states are similar to the percent level, so the shielding factor can in principle be around 100.

We demonstrate this principle experimentally by varying both technical and fundamental inhomogeneities in the localized component and observing the effect of superfluid shielding.  Our experiments begin with a nearly pure BEC in the $\left|F, m_f\right> = \left|1, -1\right>$ state, levitated against gravity by a magnetic field gradient and held in a harmonic trapping potential. Before levitation, the atom number is precisely controlled independently of the trap frequencies by varying the trap depth of a tightly confining dimple trap. In 100 ms, we ramp up a three-dimensional lattice potential with a lattice spacing of 532 nm. The vertical lattice is raised to $12\;E_r$ where $E_r$ is the recoil energy, while the transverse lattices are varied to change the densities, and therefore the interaction strengths. We then transfer a variable fraction of the atoms from the  $\left|\uparrow\right>\equiv\left|1, -1\right>$ state to the $\left|\downarrow\right>\equiv\left|2, -2\right>$ state with an RF sweep that is faster than a Bloch oscillation period. We control the ratio of the number of atoms in the two spin states by varying the intensity of the radio-frequency drive for a fixed duration of the sweep.

After state preparation, the $\left|\downarrow\right>$ atoms feel an energy gradient of $h \times 3410\;\text{Hz}$ per lattice site along the vertical direction, given by a combination of gravity and the magnetic field. This tilt is much stronger than tunneling ($\sim$$h \times 24\; \text{Hz}$), so the localization length $\sim$$\frac{J}{\Delta}a$ is much less than the lattice spacing $a$, and the state is effectively projected onto a localized number basis, creating about ten separated condensates, each with up to 3500 atoms. The remaining $\left|\uparrow\right>$ atoms, however, are still levitated against gravity and remain in a superfluid state. We allow the system to evolve for a variable time, switch off all confining potentials, and perform Stern-Gerlach separation of the spin states during ballistic expansion. An absorption image is used to measure the contrast of the resulting diffraction pattern.

\begin{figure}
\includegraphics[width=\columnwidth]{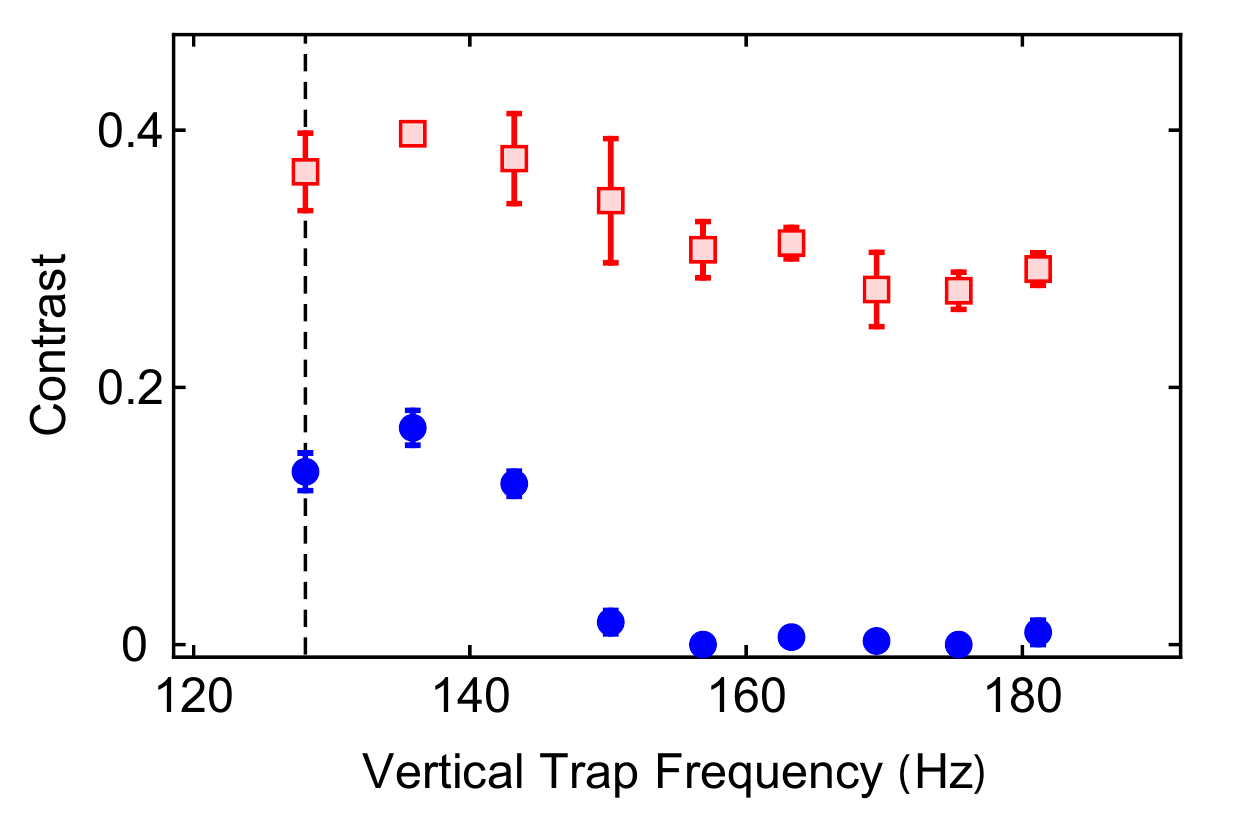}
\caption{\label{Fig6} (Color online) Superfluid shielding of external fields, applied by increasing the trap frequency.  The vertical dashed line is the initial trap frequency that corresponds to a linear chemical potential for the $\left|\downarrow\right>$ atoms. We compare the contrast after 25 Bloch oscillation cycles for unshielded (blue dots) and shielded (red squares) components. In this figure only, for technical reasons, one of the transverse lattices had a spacing of 392.5 nm.}
\end{figure}

Figure \ref{Fig1}d-f shows the central peaks of the Bragg diffraction pattern in time-of-flight of a single Bloch oscillation after ten Bloch oscillation cycles for two cases: full transfer of the ensemble to the tilted state (d), and a two-component gas with superfluid (e) and localized (f) components. The increased contrast of the superfluid peaks in the shielded system demonstrate a persistence of correlations longer than allowed by the  dephasing mechanisms affecting the system in (d). The contrast is obtained from a fit to the observed density distributions in the tilted direction and serves as the observable characterizing the phase correlations in the lattice (see Supplemental Material). An effective coherence time is obtained by fitting the decay to an exponential curve as seen in Fig. \ref{Fig2}g.  In all figures, the blue dots represent a single-component (unshielded) gas, and the red squares represent a two-component (shielded) gas. The slower decay of the shielded oscillations is clearly visible. Since the purpose of this paper is to show how extended coherence times can be achieved at strong interactions, we have intentionally increased the effect of interactions. It should be noted that longer coherence times have been observed in systems with lower densities and interactions \cite{naegerl2014,naegerl2010}.

To demonstrate superfluid shielding of common-mode external fields, curvature was intentionally added to the chemical potential by changing the external confinement after the tilt was applied.  The vertical trapping frequency was increased from $\omega_i = 2 \pi \times 128\;\mathrm{Hz}$ to a variable $\omega_f$, which adds a quadratic term to the chemical potential $\delta V_j = \frac{1}{2} m \left(\omega_f^2-\omega_i^2\right) z_j^2$, where $m$ is the $^{87}\mathrm{Rb}$ mass and $z_j$ is the position of the $j^{th}$ lattice plane. The added curvature of the chemical potential leads to dephasing and dramatically shortens the lifetime of the unshielded oscillations \cite{naegerl2010}. However, superfluid shielding can compensate for the external potential, and allow the oscillations to maintain a high contrast (Fig. \ref{Fig6}).

\begin{figure}
\includegraphics[width=\columnwidth]{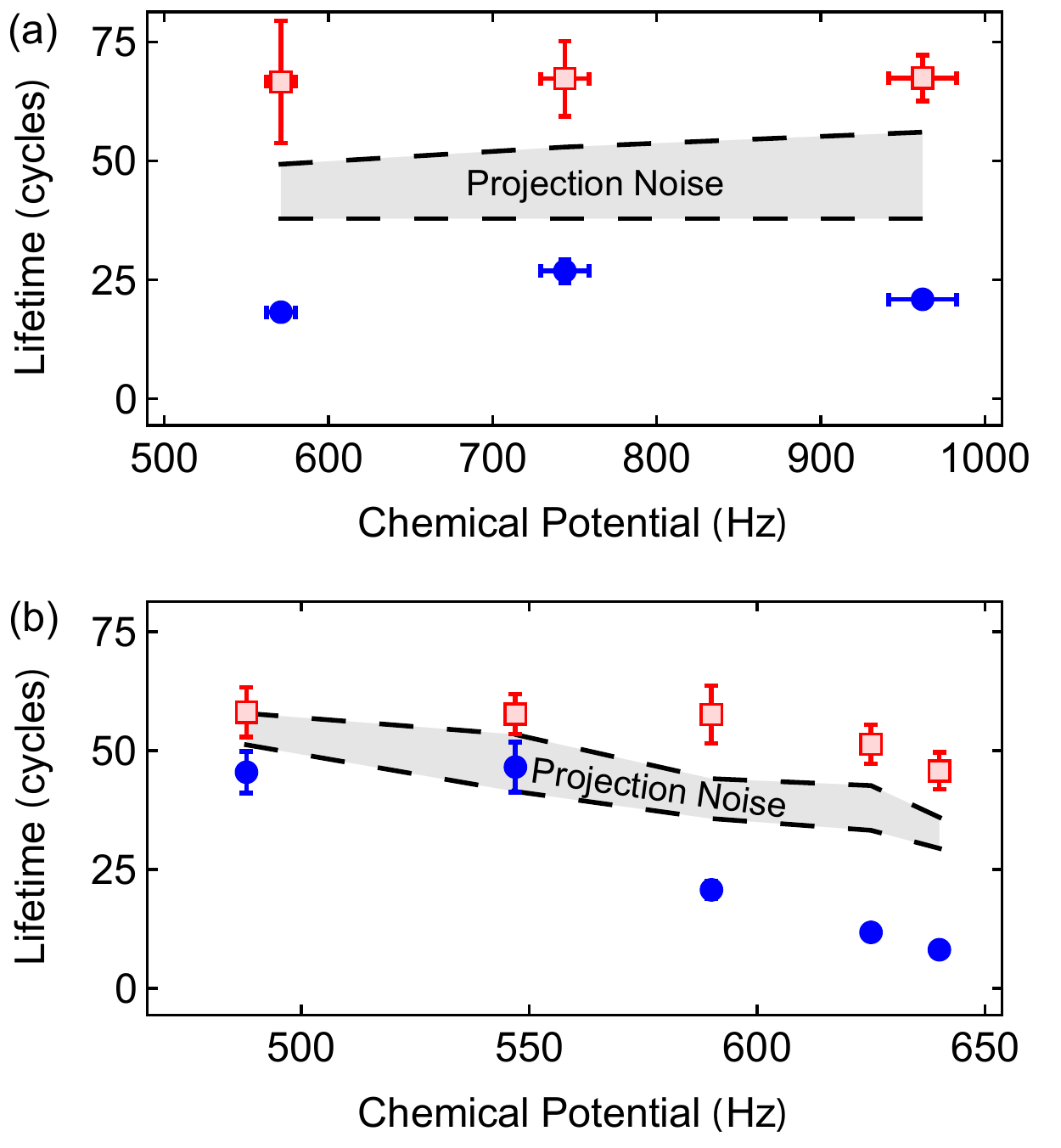}%
\caption{\label{Fig3}\label{Fig5}(Color online) Superfluid shielding for different atom numbers and densities.  Show are the exponential decay lifetimes of spin $\left|\downarrow\right>$ Bloch oscillating component for unshielded (blue dots) and shielded (red squares) cases versus chemical potential. In (a), the chemical potential is changed by varying atom number  from $\sim$$6\times10^3$ to $\sim$$2\times10^4$ while keeping the lattice depths at $10\;E_r$ in both transverse directions.  In (b), the chemical potential is varied by changing the transverse lattice depth from $4\;E_r$ to $11\;E_r$. In both plots, the dashed lines represent the projection noise limit given by two theoretical models (see text).}
\end{figure}

In the absence of perturbing external potentials, the contrast lifetime of Bloch oscillations is fundamentally limited by atomic projection noise. The dashed lines in Figs. \ref{Fig3} and \ref{Fig4} represent two models that were used to estimate the effect of projection noise. The first model (lower dashed line) assumes Poissonian shot noise in the atom number in a given plane, $\delta N_j = \sqrt{N_j}$. Finite interactions during lattice ramp-up can reduce these fluctuations by two-mode number squeezing \cite{javanainen1999} which is included in the second model (upper dashed line, see Supplemental Material).

Superfluid shielding can compensate even for this fundamental noise.  We demonstrate this principle by varying the chemical potential, changing either the atom number or the depth of the transverse lattice. For harmonic confinement in two dimensions, the Thomas-Fermi profile implies $\left|\partial \mu_j/\partial N_j\right| \propto N_j^{-1/2} U^{1/2}$, where $N_j$ is the number of atoms in a plane and $U$ is the Hubbard interaction strength for the three-dimensional lattice. Therefore, the projection noise limit does not depend on the atom number whereas the squeezed projection limit increases with increasing atom number. The observed shielded lifetimes in Fig. \ref{Fig3}a exceed the limits set by either model, and are constant to within experimental uncertainty. In Figure \ref{Fig3}b, the chemical potential was modified by varying the lattice height in the transverse, non-tilted directions, thus increasing the local density and therefore interaction strength $U$, at constant atom number. This leads to a decrease in both the shot noise and squeezed projection noise limits with increasing chemical potential. For low lattice depths, the shielded and unshielded results are consistent with the projection noise limit, but as the lattice depth increases, the unshielded lifetime decreases to below that predicted by shot noise while the shielded lifetimes consistently exceed the projection noise limit.

The difference between the unshielded lifetimes and the projection noise prediction, and the finite lifetime of the shielded sample are most likely due to technical imperfections that produce inhomogeneous chemical potentials. For the unshielded samples in Figure \ref{Fig3}, a non-adiabatic lattice ramp can lead to a curvature of the chemical potential which produces dephasing faster than that produced by projection noise. For the shielded samples, the aforementioned curvature is eliminated by superfluid shielding and dephasing is limited by the curvature of the applied external magnetic field  which we estimate to be on the order of $\sim$$100\; \mathrm{Hz}$ across the size of the sample, consistent with observed lifetimes in Figure \ref{Fig3}.

\begin{figure}
\includegraphics[width=\columnwidth]{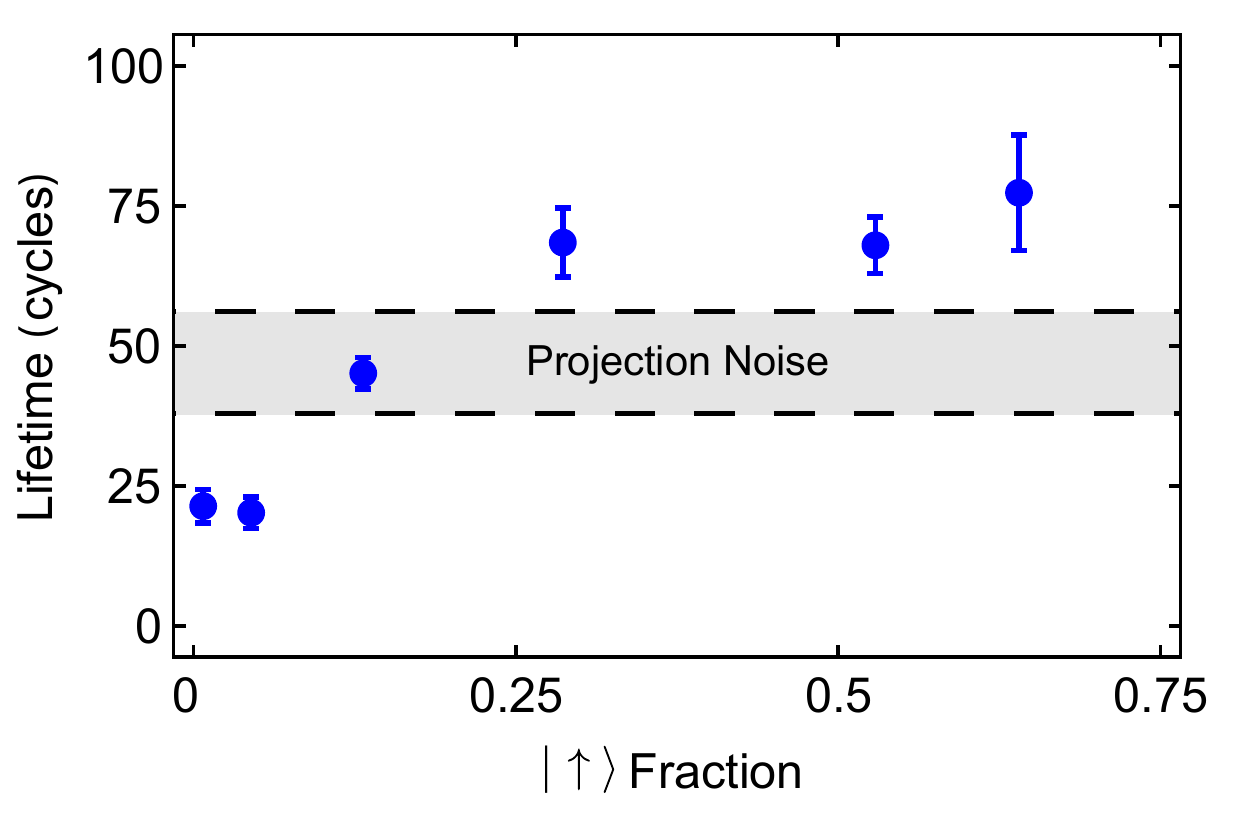}%
\caption{\label{Fig4}(Color online) Contrast lifetime versus shielding fraction. Exponential decay lifetime of spin $\left|\downarrow\right>$ Bloch oscillating component upon varying the number of atoms in spin $\left|\uparrow\right>$ state (i.e. fraction of atoms in $\left|\uparrow\right>$ state over total number of atoms) indicates that shielding is effective beyond projection noise limits once more than ~25\% atoms are in the $\left|\uparrow\right>$ state.}
\end{figure}

Finally, we demonstrate that only a small fraction of the atoms need to be in the superfluid state for the shielding effect to be robust. Figure \ref{Fig4} shows that as long as this fraction is at least 20-30\% of the atoms, full superfluid compensation is achieved. In principle, the chemical potential of the itinerant component can be as small as the residual fluctuations in the tilted component in order to provide the full shielding effect. At constant total atom number, increasing the $\left|\uparrow\right>$ fraction $f$ will reduce the number of atoms in the localized state, which decreases the signal-to-noise of the measurement by a factor of $\sqrt{1-f}$. However, as seen in Equation \ref{shieldedMu}, the shielding is a much stronger effect, and experimentally, we find an increase in lifetime by a factor of 3.2 for $f \approx 1/2$.

In conclusion, we have demonstrated superfluid shielding as a new method to increase the coherence lifetime of a quantum system through the use of interspin interactions. We have shown how superfluid shielding is a robust effect that can compensate for common-mode external fields, as well as fundamental fluctuations due to shot noise to extend the lifetime of Bloch oscillations beyond the shot noise limit. This could improve the sensitivity of force sensors based on Bloch oscillations \cite{tino2006}. In addition, although interactions are usually avoided in precision measurement, this work provides another example how interactions can enhance the performance of atomic clocks \cite{katori2008,katori2010,Ye2011} or atom interferometers \cite{ketterle2007b,ketterle2007}. 

This work focused primarily on the coherence of the localized spin, and how the itinerant component preserved it. However, the dynamics of the itinerant component is equally interesting.  Since the mean field of the localized component appears as disorder to the itinerant component, in an appropriate regime, questions of localization of the itinerant component may arise.  Finally, adding laser-assisted tunneling processes to the tilted component \cite{ketterle2013a,bloch2013,ketterle2015} enables study of an interacting two component system where only one spin is subject to a synthetic magnetic field.

\begin{acknowledgments}
We thank Alan Jamison and Ariel Sommer for helpful discussions and critical reading of the manuscript. We acknowledge Quantel Laser for their gracious support and use of an EYLSA laser for our main cooling and trapping light. W.C.C. acknowledges support of the Samsung Scholarship. This work was supported by the NSF through grant 1506369, through the Center for Ultracold Atoms, AFOSR MURI grant FA9550-14-1-0035 and ARO MURI grant W911NF-14-1-0003.
\end{acknowledgments}

\bibliography{bloch_oscillations}

\appendix
\section{Supplemental Material}
\subsection{Contrast Fitting}
Contrast is obtained from absorption images, which automatically integrate the atomic density along the imaging axis. Integrating the images along one direction gives a one-dimensional plot which is fitted with four peaks. Three of the peaks are associated with the interference peaks of condensed atoms, while the fourth is a broad background of thermal atoms. Contrast is defined as the fitted number of atoms under the condensed peaks divided by the total fitted number of atoms. This measure of contrast has a one-to-one mapping onto the coherence between lattice sites in the system. The fitting routine works reliably for contrasts below 85\%.

\subsection{Two-Mode Model of Number Squeezing}
Since we ramp up the lattice slowly before decoupling the two-dimensional slices by the sudden tilt, number fluctuations in a slice are not truly Poissonian, but rather exhibit some number squeezing. To estimate the extent of this squeezing, we use a modified two-mode model \cite{javanainen1999}, which takes into account the two-dimensional nature of the sites.

The effective Hamiltonian for large $N$ is
$$
H = -J \left(a^\dagger_l a_r + a^\dagger_r a_l \right) + \frac{2 \mu_0}{3 \sqrt{N_0}}\left(n_l^{3/2} + n_r^{3/2}\right)
$$
Here, $J$ is the tunneling energy between the left and right wells, $a_l$ and $a_r$ are the annihilation operators for the two wells, and $n_l$ and $n_r$ are the number operators. The interaction term is chosen so that $\mu = \partial E/\partial N \propto \sqrt{N}$ for both the left and right wells, as the two-dimensional Gross-Pitaevskii equation requires. The prefactor is chosen so that the chemical potential is $\mu_0$ when $N_0$ atoms are in a given well.

For a total of $2N$ atoms, the Fock state basis is  $\left|n\right> \equiv \left|N+n\right>_l \left|N-n\right>_r$. In this basis, the Hamiltonian is
$$
\begin{aligned}
H\sum_n c_n \left|n\right> = & \sum_n [ -J( \sqrt{N+n}\sqrt{N-n+1} c_{n-1} + \\
	& \sqrt{N-n}\sqrt{N+n+1} c_{n+1} ) + \\
	& \frac{2 \mu_0}{3 \sqrt{N_0}} N^{3/2} \left(2+\frac{3}{4}\left(\frac{n}{N}\right)^2\right) c_n] \left|n\right>,
\end{aligned}
$$
dropping terms of order $O\left(\left(\frac{n}{N}\right)^4\right)$ and higher. Following \cite{javanainen1999}, we assume that $c_{n\pm1}$ can be written as $c_{n\pm1} \equiv C(n\pm1) = C(n) \pm C'(n) + \frac{1}{2} C''(n) + \ldots$ and expand the coefficients $C(n),\;C'(n),\;C''(n)$ in powers of $1/N$. After dropping constant terms and orders higher than $1/N$, we find the eigenvalue equation
$$
\epsilon C(n) = \left(-J N \partial_n^2 + \left(\frac{J}{N} + \frac{3}{4} \frac{2 \mu_0}{3 \sqrt{N_0}} \frac{1}{N^{1/2}}\right) n^2 \right) C(n)
$$
which is equivalent to the Schroedinger equation for the simple harmonic oscillator, and has a Gaussian ground state with a sub-Poissonian width
$$
\delta n^2 = N \left(\frac{J}{J+\frac{\mu_0 \sqrt{N/N_0}}{2}}\right)^{1/2}.
$$

\subsection{Projection Noise Limit Simulations}
The three-dimensional system is simulated using the discrete Gross-Pitaevskii equation along the lattice direction with a discrete Thomas-Fermi profile along the transverse directions.  This provides the average occupation number in every two-dimensional slice of the system. Since we apply the tilt suddenly, these states are then projected onto states with a definite number, randomly selected from either a Poissonian distribution, or a Gaussian distribution with a width given by the squeezed number fluctuations. The chemical potential on site $j$ is given by the two-dimensional Gross-Pitaevskii equation, which determines the time evolution of the phase in a given two-dimensional slice $\left|\psi_j(t)\right> = \exp[-i \mu_j t/\hbar] \left|\psi_j(0)\right>$.

We time evolve the system, assuming that the only relevant dynamics are the evolution of the phases, and then take a Fourier transformation to simulate time-of-flight expansion. Twenty simulated images are averaged and the contrast is obtained using the same fitting method as for our experimental data. Because our simulations start at perfect contrast, while the experiments do not, and because our fitting routine is unreliable at low incoherent fraction, we time-evolve the system until its contrast matches the initial experimental contrast before measuring the decay time.

\end{document}